\documentclass[%
 reprint,
 superscriptaddress,
 twocolumn,
 showpacs,
 preprintnumbers,
 amsmath,amssymb,
 aps,
]{revtex4-1}

\usepackage{graphicx}
\usepackage{dcolumn}
\usepackage{bm}
\usepackage{slashed}
\usepackage{braket}
\usepackage{multirow}

\def \nobreakseq {\nobreak \hskip 0pt \hbox}

\newcommand{\ca}{$^{48}{\rm Ca}$}
\newcommand{\bbtwo}{$2\nu\beta\beta$}
\newcommand{\bbzero}{$0\nu\beta\beta$}

\begin{document}

\title{Measurement of the double-beta decay half-life and search for the
  neutrinoless double-beta decay of $^{48}{\rm Ca}$ with the NEMO-3 detector}

\setcounter{footnote}{2}

\author{R.~Arnold}
\affiliation{IPHC, ULP, CNRS/IN2P3\nobreakseq{,} F-67037 Strasbourg, France}
\author{C.~Augier} 
\affiliation{LAL, Universit\'{e} Paris-Sud\nobreakseq{,} CNRS/IN2P3\nobreakseq{,}
  Universit\'{e} Paris-Saclay\nobreakseq{,} F-91405 Orsay\nobreakseq{,} France}
\author{A.M.~Bakalyarov}
\affiliation{NRC ``Kurchatov Institute", KI, 123182 Moscow, Russia}
\author{J.D.~Baker}
\thanks{Deceased}
\affiliation{Idaho National Laboratory\nobreakseq{,} Idaho Falls, ID 83415,
  U.S.A.}
\author{A.S.~Barabash}
\affiliation{NRC ``Kurchatov Institute", ITEP, 117218 Moscow, Russia}
\author{A.~Basharina-Freshville} 
\affiliation{UCL, London WC1E 6BT\nobreakseq{,} United Kingdom}
\author{S.~Blondel} 
\affiliation{LAL, Universit\'{e} Paris-Sud\nobreakseq{,} CNRS/IN2P3\nobreakseq{,}
  Universit\'{e} Paris-Saclay\nobreakseq{,} F-91405 Orsay\nobreakseq{,} France}
\author{S.~Blot}
\affiliation{University of Manchester\nobreakseq{,} Manchester M13
  9PL\nobreakseq{,}~United Kingdom}
\author{M.~Bongrand} 
\affiliation{LAL, Universit\'{e} Paris-Sud\nobreakseq{,} CNRS/IN2P3\nobreakseq{,}
  Universit\'{e} Paris-Saclay\nobreakseq{,} F-91405 Orsay\nobreakseq{,} France}
\author{V.~Brudanin} 
\affiliation{JINR, 141980 Dubna, Russia}
\affiliation{National Research Nuclear University MEPhI, 115409 Moscow, Russia}
\author{J.~Busto} 
\affiliation{CPPM, Universit\'e de Marseille\nobreakseq{,}
  CNRS/IN2P3\nobreakseq{,} F-13288 Marseille\nobreakseq{,} France}
\author{A.J.~Caffrey}
\affiliation{Idaho National Laboratory\nobreakseq{,} Idaho Falls, ID 83415,
  U.S.A.}
\author{S. Calvez}
\affiliation{LAL, Universit\'{e} Paris-Sud\nobreakseq{,} CNRS/IN2P3\nobreakseq{,}
  Universit\'{e} Paris-Saclay\nobreakseq{,} F-91405 Orsay\nobreakseq{,} France}
\author{M.~Cascella} 
\affiliation{UCL, London WC1E 6BT\nobreakseq{,} United Kingdom}
\author{C.~Cerna} 
\affiliation{CENBG\nobreakseq{,} Universit\'e de Bordeaux\nobreakseq{,}
  CNRS/IN2P3\nobreakseq{,} F-33175 Gradignan\nobreakseq{,} France}
\author{J.P.~Cesar}
\affiliation{University of Texas at Austin\nobreakseq{,}
  Austin\nobreakseq{,} TX 78712\nobreakseq{,}~U.S.A.}
\author{A.~Chapon} 
\affiliation{LPC Caen\nobreakseq{,} ENSICAEN\nobreakseq{,} Universit\'e de
Caen\nobreakseq{,} CNRS/IN2P3\nobreakseq{,} F-14050 Caen\nobreakseq{,} France}
\author{E.~Chauveau} 
\affiliation{University of Manchester\nobreakseq{,} Manchester M13
  9PL\nobreakseq{,}~United Kingdom}
\author{A.~Chopra} 
\affiliation{UCL, London WC1E 6BT\nobreakseq{,} United Kingdom}
\author{D.~Duchesneau} 
\affiliation{LAPP, Universit\'e de Savoie\nobreakseq{,}
CNRS/IN2P3\nobreakseq{,} F-74941 Annecy-le-Vieux\nobreakseq{,} France}
\author{D.~Durand} 
\affiliation{LPC Caen\nobreakseq{,} ENSICAEN\nobreakseq{,} Universit\'e de
Caen\nobreakseq{,} CNRS/IN2P3\nobreakseq{,} F-14050 Caen\nobreakseq{,} France}
\author{V.~Egorov}
\affiliation{JINR, 141980 Dubna, Russia}
\author{G.~Eurin} 
\affiliation{LAL, Universit\'{e} Paris-Sud\nobreakseq{,} CNRS/IN2P3\nobreakseq{,}
  Universit\'{e} Paris-Saclay\nobreakseq{,} F-91405 Orsay\nobreakseq{,} France}
\affiliation{UCL, London WC1E 6BT\nobreakseq{,} United Kingdom}
\author{J.J.~Evans} 
\affiliation{University of Manchester\nobreakseq{,} Manchester M13
  9PL\nobreakseq{,}~United Kingdom}
\author{L.~Fajt} 
\affiliation{Institute of Experimental and Applied Physics\nobreakseq{,} Czech
  Technical University in Prague\nobreakseq{,} CZ-12800 Prague\nobreakseq{,}
  Czech Republic}
\author{D.~Filosofov} 
\affiliation{JINR, 141980 Dubna, Russia}
\author{R.~Flack} 
\affiliation{UCL, London WC1E 6BT\nobreakseq{,} United Kingdom}
\author{X.~Garrido} 
\affiliation{LAL, Universit\'{e} Paris-Sud\nobreakseq{,} CNRS/IN2P3\nobreakseq{,}
  Universit\'{e} Paris-Saclay\nobreakseq{,} F-91405 Orsay\nobreakseq{,} France}
\author{H.~G\'omez} 
\affiliation{LAL, Universit\'{e} Paris-Sud\nobreakseq{,} CNRS/IN2P3\nobreakseq{,}
  Universit\'{e} Paris-Saclay\nobreakseq{,} F-91405 Orsay\nobreakseq{,} France}
\author{B.~Guillon} 
\affiliation{LPC Caen\nobreakseq{,} ENSICAEN\nobreakseq{,} Universit\'e de
Caen\nobreakseq{,} CNRS/IN2P3\nobreakseq{,} F-14050 Caen\nobreakseq{,} France}
\author{P.~Guzowski} 
\affiliation{University of Manchester\nobreakseq{,} Manchester M13
  9PL\nobreakseq{,}~United Kingdom}
\author{R.~Hod\'{a}k} 
\affiliation{Institute of Experimental and Applied Physics\nobreakseq{,} Czech
Technical University in Prague\nobreakseq{,} CZ-12800
Prague\nobreakseq{,} Czech Republic}
\author{A.~Huber} 
\affiliation{CENBG\nobreakseq{,} Universit\'e de Bordeaux\nobreakseq{,}
  CNRS/IN2P3\nobreakseq{,} F-33175 Gradignan\nobreakseq{,} France}
\author{P.~Hubert} 
\affiliation{CENBG\nobreakseq{,} Universit\'e de Bordeaux\nobreakseq{,}
  CNRS/IN2P3\nobreakseq{,} F-33175 Gradignan\nobreakseq{,} France}
\author{C.~Hugon}
\affiliation{CENBG\nobreakseq{,} Universit\'e de Bordeaux\nobreakseq{,}
  CNRS/IN2P3\nobreakseq{,} F-33175 Gradignan\nobreakseq{,} France}
\author{S.~Jullian} 
\affiliation{LAL, Universit\'{e} Paris-Sud\nobreakseq{,} CNRS/IN2P3\nobreakseq{,}
  Universit\'{e} Paris-Saclay\nobreakseq{,} F-91405 Orsay\nobreakseq{,} France}
\author{A.~Klimenko} 
\affiliation{JINR, 141980 Dubna, Russia}
\author{O.~Kochetov} 
\affiliation{JINR, 141980 Dubna, Russia}
\author{S.I.~Konovalov} 
\affiliation{NRC ``Kurchatov Institute", ITEP, 117218 Moscow, Russia}
\author{V.~Kovalenko}
\affiliation{JINR, 141980 Dubna, Russia}
\author{D.~Lalanne} 
\affiliation{LAL, Universit\'{e} Paris-Sud\nobreakseq{,} CNRS/IN2P3\nobreakseq{,}
  Universit\'{e} Paris-Saclay\nobreakseq{,} F-91405 Orsay\nobreakseq{,} France}
\author{K.~Lang} 
\affiliation{University of Texas at Austin\nobreakseq{,}
  Austin\nobreakseq{,} TX 78712\nobreakseq{,}~U.S.A.}
\author{V.I.~Lebedev}
\affiliation{NRC ``Kurchatov Institute", KI, 123182 Moscow, Russia}
\author{Y.~Lemi\`ere} 
\affiliation{LPC Caen\nobreakseq{,} ENSICAEN\nobreakseq{,} Universit\'e de
Caen\nobreakseq{,} CNRS/IN2P3\nobreakseq{,} F-14050 Caen\nobreakseq{,} France}
\author{T.~Le~Noblet} 
\affiliation{LAPP, Universit\'e de Savoie\nobreakseq{,} CNRS/IN2P3\nobreakseq{,}
  F-74941 Annecy-le-Vieux\nobreakseq{,} France}
\author{Z.~Liptak} 
\affiliation{University of Texas at Austin\nobreakseq{,}
  Austin\nobreakseq{,} TX 78712\nobreakseq{,}~U.S.A.}
\author{X.~R.~Liu} 
\affiliation{UCL, London WC1E 6BT\nobreakseq{,} United Kingdom}  
\author{P.~Loaiza} 
\affiliation{LAL, Universit\'{e} Paris-Sud\nobreakseq{,} CNRS/IN2P3\nobreakseq{,}
  Universit\'{e} Paris-Saclay\nobreakseq{,} F-91405 Orsay\nobreakseq{,} France}
\author{G.~Lutter} 
\affiliation{CENBG\nobreakseq{,} Universit\'e de Bordeaux\nobreakseq{,}
  CNRS/IN2P3\nobreakseq{,} F-33175 Gradignan\nobreakseq{,} France}
\author{F.~Mamedov}
\affiliation{Institute of Experimental and Applied Physics\nobreakseq{,} Czech
Technical University in Prague\nobreakseq{,} CZ-12800
Prague\nobreakseq{,} Czech Republic}
\author{C.~Marquet} 
\affiliation{CENBG\nobreakseq{,} Universit\'e de Bordeaux\nobreakseq{,}
  CNRS/IN2P3\nobreakseq{,} F-33175 Gradignan\nobreakseq{,} France}
\author{F.~Mauger} 
\affiliation{LPC Caen\nobreakseq{,} ENSICAEN\nobreakseq{,} Universit\'e de
Caen\nobreakseq{,} CNRS/IN2P3\nobreakseq{,} F-14050 Caen\nobreakseq{,} France}
\author{B.~Morgan} 
\affiliation{University of Warwick\nobreakseq{,} Coventry CV4
7AL\nobreakseq{,} United Kingdom}
\author{J.~Mott} 
\affiliation{UCL, London WC1E 6BT\nobreakseq{,} United Kingdom}
\author{I.~Nemchenok} 
\affiliation{JINR, 141980 Dubna, Russia}
\author{M.~Nomachi} 
\affiliation{Osaka University\nobreakseq{,} 1-1 Machikaneyama
Toyonaka\nobreakseq{,} Osaka 560-0043\nobreakseq{,} Japan}
\author{F.~Nova} 
\affiliation{University of Texas at Austin\nobreakseq{,}
  Austin\nobreakseq{,} TX 78712\nobreakseq{,}~U.S.A.}
\author{F.~Nowacki} 
\affiliation{IPHC, ULP, CNRS/IN2P3\nobreakseq{,} F-67037 Strasbourg, France}
\author{H.~Ohsumi} 
\affiliation{Saga University\nobreakseq{,} Saga 840-8502\nobreakseq{,}
  Japan}
\author{R.B.~Pahlka}
\affiliation{University of Texas at Austin\nobreakseq{,}
  Austin\nobreakseq{,} TX 78712\nobreakseq{,}~U.S.A.}
\author{F.~Perrot} 
\affiliation{CENBG\nobreakseq{,} Universit\'e de Bordeaux\nobreakseq{,}
  CNRS/IN2P3\nobreakseq{,} F-33175 Gradignan\nobreakseq{,} France}
\author{F.~Piquemal} 
\affiliation{CENBG\nobreakseq{,} Universit\'e de Bordeaux\nobreakseq{,}
  CNRS/IN2P3\nobreakseq{,} F-33175 Gradignan\nobreakseq{,} France}
\affiliation{Laboratoire Souterrain de Modane\nobreakseq{,} F-73500
Modane\nobreakseq{,} France}
\author{P.~Povinec}
\affiliation{FMFI,~Comenius~Univ.\nobreakseq{,}
SK-842~48~Bratislava\nobreakseq{,}~Slovakia}
\author{P.~P\v{r}idal} 
\affiliation{Institute of Experimental and Applied Physics\nobreakseq{,} Czech
  Technical University in Prague\nobreakseq{,} CZ-12800
  Prague\nobreakseq{,} Czech Republic}
\author{Y.A.~Ramachers} 
\affiliation{University of Warwick\nobreakseq{,} Coventry CV4
7AL\nobreakseq{,} United Kingdom}
\author{A.~Remoto}
\affiliation{LAPP, Universit\'e de Savoie\nobreakseq{,}
CNRS/IN2P3\nobreakseq{,} F-74941 Annecy-le-Vieux\nobreakseq{,} France}
\author{J.L.~Reyss} 
\affiliation{LSCE\nobreakseq{,} CNRS\nobreakseq{,} F-91190
  Gif-sur-Yvette\nobreakseq{,} France}
\author{B.~Richards} 
\affiliation{UCL, London WC1E 6BT\nobreakseq{,} United Kingdom}
\author{C.L.~Riddle} 
\affiliation{Idaho National Laboratory\nobreakseq{,} Idaho Falls, ID 83415,
  U.S.A.}
\author{E.~Rukhadze} 
\affiliation{Institute of Experimental and Applied Physics\nobreakseq{,} Czech
Technical University in Prague\nobreakseq{,} CZ-12800
Prague\nobreakseq{,} Czech Republic}
\author{N.I.~Rukhadze}
\affiliation{JINR, 141980 Dubna, Russia}
\author{R.~Saakyan} 
\affiliation{UCL, London WC1E 6BT\nobreakseq{,} United Kingdom}
\author{R.~Salazar} 
\affiliation{University of Texas at Austin\nobreakseq{,}
  Austin\nobreakseq{,} TX 78712\nobreakseq{,}~U.S.A.}
\author{X.~Sarazin} 
\affiliation{LAL, Universit\'{e} Paris-Sud\nobreakseq{,} CNRS/IN2P3\nobreakseq{,}
  Universit\'{e} Paris-Saclay\nobreakseq{,} F-91405 Orsay\nobreakseq{,} France}
\author{Yu.~Shitov} 
\affiliation{JINR, 141980 Dubna, Russia}
\affiliation{Imperial College London\nobreakseq{,} London SW7
2AZ\nobreakseq{,} United Kingdom}
\author{L.~Simard} 
\affiliation{LAL, Universit\'{e} Paris-Sud\nobreakseq{,} CNRS/IN2P3\nobreakseq{,}
  Universit\'{e} Paris-Saclay\nobreakseq{,} F-91405 Orsay\nobreakseq{,} France}
\affiliation{Institut Universitaire de France\nobreakseq{,} F-75005
  Paris\nobreakseq{,} France}
\author{F.~\v{S}imkovic} 
\affiliation{FMFI,~Comenius~Univ.\nobreakseq{,}
  SK-842~48~Bratislava\nobreakseq{,}~Slovakia}
\author{A.~Smetana}
\affiliation{Institute of Experimental and Applied Physics\nobreakseq{,} Czech
Technical University in Prague\nobreakseq{,} CZ-12800
Prague\nobreakseq{,} Czech Republic}
\author{K.~Smolek} 
\affiliation{Institute of Experimental and Applied Physics\nobreakseq{,} Czech
Technical University in Prague\nobreakseq{,} CZ-12800
Prague\nobreakseq{,} Czech Republic}
\author{A.~Smolnikov} 
\affiliation{JINR, 141980 Dubna, Russia}
\author{S.~S\"oldner-Rembold}
\affiliation{University of Manchester\nobreakseq{,} Manchester M13
  9PL\nobreakseq{,}~United Kingdom}
\author{B.~Soul\'e}
\affiliation{CENBG\nobreakseq{,} Universit\'e de Bordeaux\nobreakseq{,}
  CNRS/IN2P3\nobreakseq{,} F-33175 Gradignan\nobreakseq{,} France}
\author{I.~\v{S}tekl} 
\affiliation{Institute of Experimental and Applied Physics\nobreakseq{,} Czech
  Technical University in Prague\nobreakseq{,} CZ-12800
  Prague\nobreakseq{,} Czech Republic}
\author{J.~Suhonen} 
\affiliation{Jyv\"askyl\"a University\nobreakseq{,} FIN-40351
  Jyv\"askyl\"a\nobreakseq{,} Finland}
\author{C.S.~Sutton} 
\affiliation{MHC\nobreakseq{,} South Hadley\nobreakseq{,} Massachusetts
  01075\nobreakseq{,} U.S.A.}
\author{G.~Szklarz}
\affiliation{LAL, Universit\'{e} Paris-Sud\nobreakseq{,} CNRS/IN2P3\nobreakseq{,}
  Universit\'{e} Paris-Saclay\nobreakseq{,} F-91405 Orsay\nobreakseq{,} France}
\author{J.~Thomas} 
\affiliation{UCL, London WC1E 6BT\nobreakseq{,} United Kingdom}
\author{V.~Timkin} 
\affiliation{JINR, 141980 Dubna, Russia}
\author{S.~Torre} 
\affiliation{UCL, London WC1E 6BT\nobreakseq{,} United Kingdom}
\author{Vl.I.~Tretyak} 
\affiliation{Institute for Nuclear Research\nobreakseq{,} MSP 03680\nobreakseq{,}
  Kyiv\nobreakseq{,} Ukraine}
\author{V.I.~Tretyak}
\affiliation{JINR, 141980 Dubna, Russia}
\author{V.I.~Umatov} 
\affiliation{NRC ``Kurchatov Institute", ITEP, 117218 Moscow, Russia}
\author{I.~Vanushin} 
\affiliation{NRC ``Kurchatov Institute", ITEP, 117218 Moscow, Russia}
\author{C.~Vilela} 
\affiliation{UCL, London WC1E 6BT\nobreakseq{,} United Kingdom}
\author{V.~Vorobel} 
\affiliation{Charles University in Prague\nobreakseq{,} Faculty of Mathematics
and Physics\nobreakseq{,} CZ-12116 Prague\nobreakseq{,} Czech Republic}
\author{D.~Waters} 
\affiliation{UCL, London WC1E 6BT\nobreakseq{,} United Kingdom}
\author{S.V.~Zhukov}
\affiliation{NRC ``Kurchatov Institute", KI, 123182 Moscow, Russia}
\author{A.~\v{Z}ukauskas}
\affiliation{Charles University in Prague\nobreakseq{,} Faculty of Mathematics
and Physics\nobreakseq{,} CZ-12116 Prague\nobreakseq{,} Czech Republic}
\collaboration{NEMO-3 Collaboration}
\noaffiliation

\date{\today}

\begin{abstract}
The NEMO-3 experiment at the Modane Underground Laboratory investigates the
double-beta decay of \ca{}. Using $5.25$\,yr of data recorded with a
$6.99\,{\rm g}$ sample of \ca{}, approximately $150$ double-beta decay
candidate events are selected with a signal-to-background ratio greater
than $3$. The half-life for the two-neutrino double-beta decay of \ca{} 
is measured to be \mbox{$T^{2\nu}_{1/2}\,=\,[6.4\, ^{+0.7}_{-0.6}{\rm (stat.)}
    \, ^{+1.2}_{-0.9}{\rm (syst.)}] \times 10^{19}\,{\rm yr}$}. A search for
neutrinoless double-beta decay of \ca{} yields a null result and a
corresponding lower limit on the half-life is found to be $T^{0\nu}_{1/2} > 2.0
\times 10^{22}\,{\rm yr}$ at $90\%$ confidence level, translating into an upper
limit on the effective Majorana neutrino mass of $\braket{m_{\beta\beta}} < 6.0
- 26$\,${\rm eV}$, with the range reflecting different nuclear matrix element
calculations. Limits are also set on models involving Majoron emission and
right-handed currents.
\end{abstract}

\pacs{23.40.-s; 14.60.Pq}

\maketitle

\section{\label{sec:introduction}Introduction}

Neutrinoless double-beta decay (\bbzero{}) is the only feasible way of
experimentally determining the Majorana or Dirac nature of light neutrinos. In
the event that neutrinos are Majorana particles, this process may provide one of
the most promising methods for measuring their absolute mass as well as the
possible presence of additional CP-violating phases (see for
example~\cite{0vBB_review}).  The corresponding two-neutrino decay mode
(\bbtwo{}) does not violate lepton number conservation and is allowed in the
standard model. This mode has been observed for several isotopes and provides
valuable inputs to nuclear matrix element (NME) calculations~\cite{2vBB_review,
  barabash_2015}. However no clear evidence for \bbzero{} has been established
to date, with the best half-life limits in the range $10^{24-25}$\,${\rm
  yr}$~\cite{cuoricino, exo200,kamland_zen,gerda,nemo3_mo100}.

\ca{} is a particularly interesting nucleus with which to search for
neutrinoless double-beta decay since it has the highest kinetic energy
release \mbox{$Q_{\beta\beta} = (4267.98 \pm 0.32)$\,${\rm keV}$} of any known
double-beta decaying isotope~\cite{ca48_Qbb}. This $Q_{\beta\beta}$ value is
significantly above the bulk of naturally occurring radioactive backgrounds and,
moreover, ensures a favorable phase space which enhances the \bbzero{} decay
rate. However \ca{} has a low natural abundance of $0.187\%$ and is difficult to
enrich. Recent calculations also indicate a relative suppression of \bbzero{}
NMEs for \ca{} compared to other isotopes~\cite{nme_ism_ca48, horoi_2013,
  nme_edf_ca48, nme_qrpa_ca48, suhonen_1993,simkovic_2013, nme_ibm_ca48_2015,
  iwata_2016}.

The \bbtwo{} decay mode of \ca{} was first discovered in the Hoover Dam TPC
experiment, employing approximately $14$\,${\rm g}$ of
isotope~\cite{balysh_1996}.  With a sample of approximately $100$ events, the
half-life was measured to be $T_{1/2}^{2\nu} = [4.3^{+2.4}_{-1.1} {\rm (stat)}
  \pm 1.4 {\rm (syst)}] \times 10^{19}$\,${\rm yr}$.  A subsequent measurement
in the TGV planar germanium array experiment using $1$\,${\rm g}$ of isotope
yielded $T_{1/2}^{2\nu} = 4.2^{+3.3}_{-1.3} \times 10^{19}$\,${\rm yr}$ based on
only five events in the region of interest~\cite{brudanin_2000}.  Both of these
measurements are in agreement with the shell-model prediction of
\mbox{$T_{1/2}^{2\nu} = 3.7 \times 10^{19}$\,${\rm yr}$~\cite{caurier_1994}}.
Shell-model calculations are expected to be the most reliable for
\ca{}. Nevertheless recent predictions for the two-neutrino half-life span a
range of $(2.0-4.5) \times 10^{19}$\,${\rm yr}$~\cite{caurier_2012, horoi_2007,
  zhao_1990, iwata_2016}.  Other NME schemes are less well suited to this
isotope, although there is a quasiparticle random-phase approximation (QRPA)
prediction of $4.7 \times 10^{19}$\,${\rm yr}$~\cite{raduta_2011}.

Searches for the \bbzero{} decay of \ca{} began 60 years
ago~\cite{mccarthy_1955}.  The main experimental technique that has been
employed is the use of scintillating ${\rm CaF_{2}}$
crystals~\cite{mateosian_1966, beijing, elegant_2008}. Such detectors have good
energy resolution and detection efficiency in order to search for a line at
$Q_{\beta\beta}$, while backgrounds generally prevent a straightforward
measurement of the \bbtwo{} rate. No signals for \bbzero{} have been found, with
the best limit currently set by the \mbox{ELEGANT~VI} experiment using
$6.6$\,${\rm kg}$ of ${\rm CaF_{2}}$ ($7.6$\,${\rm g}$ of \ca{}) at
$T_{1/2}^{0\nu} > 5.8 \times 10^{22}$\,${\rm yr}$ at
$90\%$~C.L.~\cite{elegant_2008}.

\section{\label{sec:NEMO-3}NEMO-3 Detector}

NEMO-3 is a tracking calorimeter detector, hosting several double-beta
decaying isotopes in thin source foils arranged in a cylindrical
geometry~\cite{nemo3_detector}. Electrons pass through $50$\,${\rm cm}$ wide
wire chambers on each side of the source foils, containing in total 6180 Geiger
cells operating in a gas mixture comprising helium with $4\%$ ethanol quencher,
$1\%$ argon and $0.15\%$ water vapor. Surrounding the tracker is the
calorimeter, consisting of 1940 plastic scintillators coupled to low
radioactivity photomultipliers. A copper coil mounted outside the calorimeter
generates a $25$\,${\rm G}$ solenoidal magnetic field and beyond that 165 tons
of iron, in addition to borated water, paraffin and wood, are used to shield the
inner detector from external radioactivity. Approximately $95\%$ of $1$\,${\rm
  MeV}$ positrons are rejected by determining the track curvature, and
$1$\,${\rm MeV}$ electrons are measured in the calorimeter with an energy
resolution (FWHM) ranging from $14.1\%$ to $17.7\%$ and a timing resolution
$\sigma_{t}\approx 250$\,${\rm ps}$. A cylindrical coordinate system $(R,\phi,
Z)$ is used, with the $Z$ axis pointing upwards, parallel to the source foil and
tracker wires.

\begin{figure}[tb]
  \centering
  \includegraphics[width=\columnwidth]{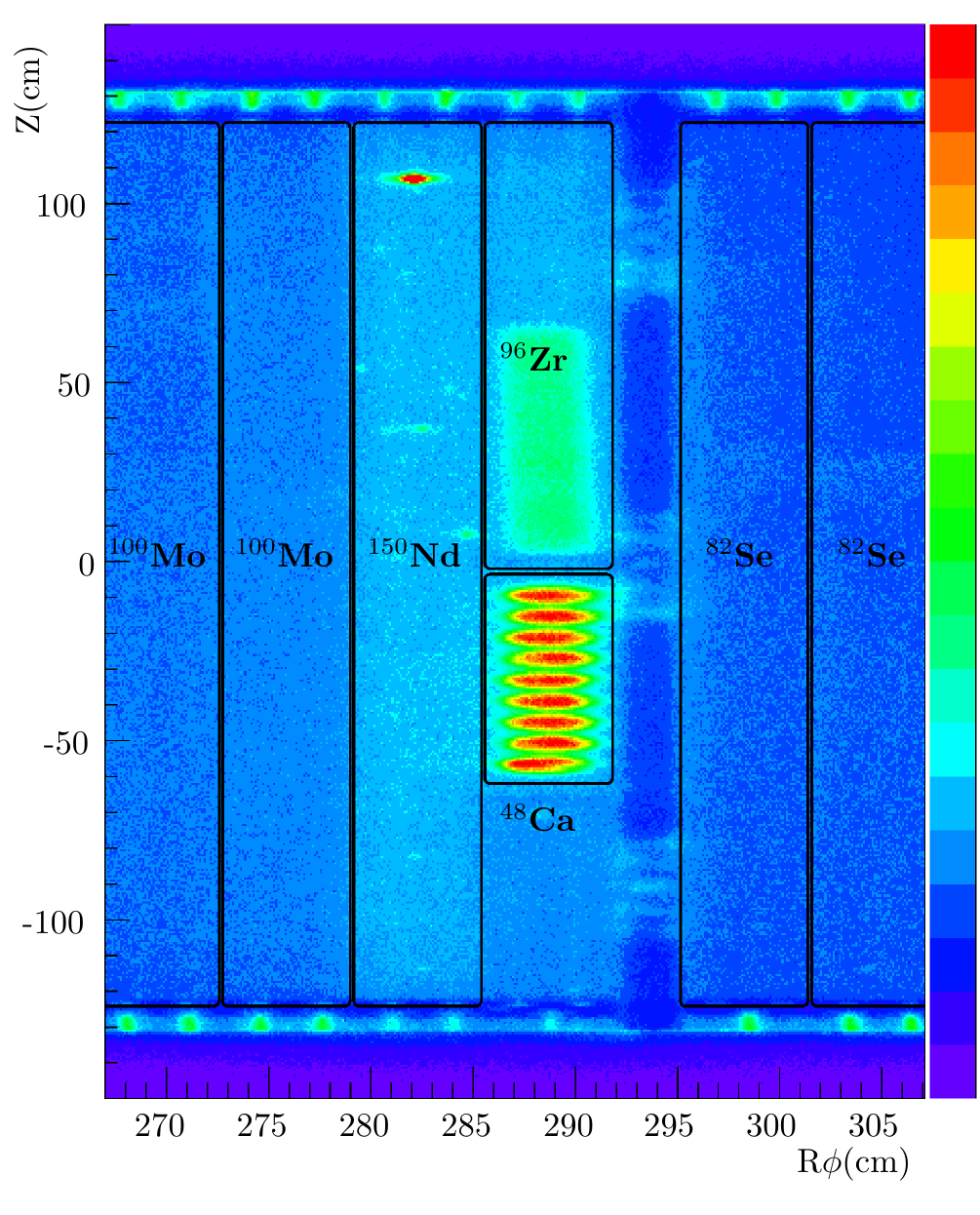}
  \caption{\label{fig:ca_source_image} An image of the NEMO-3 source foils in
    the vicinity of the \ca{} source, showing the rate of single-electron events
    (arbitrary scale) from the nine ${\rm CaF_{2}}$ disks mounted beneath the
    $^{96}{\rm Zr}$ source and next to the $^{150}{\rm Nd}$ source foil. The
    region below the \ca{} source is populated with Mylar foils of the type
    used in the source construction. The more distant $^{82}{\rm Se}$ and
    $^{100}{\rm Mo}$ sources do not contribute to backgrounds in the analysis of
    \ca{}. A copper tube forming part of the calibration system lies between the
    $^{48}{\rm Ca}/^{96}{\rm Zr}$ sources and the $^{82}{\rm Se}$ foil.}
\end{figure}

NEMO-3 hosts $6.91$\,${\rm kg}$ of $^{100}{\rm Mo}$ and $0.93$\,${\rm kg}$ of
$^{82}{\rm Se}$, as well as smaller amounts of $^{96}{\rm Zr}$, $^{116}{\rm
  Cd}$, $^{130}{\rm Te}$ and $^{150}{\rm Nd}$.  In all cases the experiment has
made the most precise measurement of the $2\nu\beta\beta$
half-life~\cite{nemo3_Nd150_2009, nemo3_Zr96_2010, nemo3_Te130_2011,
  2vBB_review}. The \ca{} source in NEMO-3 consists of nine disks of ${\rm
  CaF_{2}}$ powder contained in thin ($\approx 10$\,$\mu{\rm m}$) Teflon and
polyethylene envelopes. Mylar foils, approximately $20\,\mu{\rm m}$ thick,
sandwich the disks on either side in order to hold them in place. The disks have
a diameter of $46$\,${\rm mm}$ and contain in total $17.5\,{\rm g}$ of ${\rm
  CaF_{2}}$. With an enrichment fraction of $(73.2\pm 1.6)\%$, this corresponds
to $6.99\,{\rm g}$ of \ca{}. Following the dismantling of the detector, more
accurate measurements of the disks' diameter and orientation within the detector
were made which resulted in an updated description of the source geometry.
Figure~\ref{fig:ca_source_image} shows the vicinity
of the \ca{} sources, imaged in single-electron events.

NEMO-3 ran from February 2003 until January 2011. A standard good-run
requirement selects data-taking periods during which the detector was stable and
the calorimeter was well calibrated. The resulting live time is $5.25$\,${\rm
  yr}$, corresponding to a \ca{} exposure of $36.7$\,${\rm g\cdot yr}$.

\section{\label{sec:selection}Event Reconstruction and Selection}


Readout of NEMO-3 is triggered by a calorimeter energy deposit of at least
$150$\,${\rm keV}$ in temporal and geometrical coincidence with a number of hits
in the tracking detector, which has a negligible inefficiency for fiducial \ca{}
two-electron ($\beta\beta$) events. Offline, a tracking algorithm fits helices
to clusters of Geiger cell hits, and performs additional fits to localize the
origin of each track on the source foil and the entry point on the front face of
the calorimeter.  Isolated calorimeter hits that have no energy deposits in
neighboring calorimeter cells and that are matched to reconstructed tracks have
their energies corrected for the track impact position and are stored as
electron candidates. For all event topologies containing a single electron, the
sign of the track curvature is required to be consistent with a negatively
charged particle.


Selected $\beta\beta$ events must contain two electrons each with
\mbox{$E_{e}>400$\,${\rm keV}$}. The two tracks must intersect the \ca{} source
strip and originate from the same point within tolerances
$|\Delta_{XY}|<10$\,${\rm cm}$ and $|\Delta_{Z}|<15$\,${\rm cm}$. In addition
there must be no delayed \mbox{$\alpha$-particle} tracks, as defined in
Sec.~\ref{sec:backgrounds}, in the event.  The calorimeter times of the two
electrons must be consistent with a common origin on the source foil and
inconsistent with a crossing particle originating from outside the detector. No
requirement is made on the number of additional calorimeter hits that may be
present in the event.


Backgrounds are largely constrained in independent control samples. Common
minimum electron and photon energy cuts of $400$\,${\rm keV}$ ensure good
agreement between data and Monte Carlo simulations across all the
samples. Events containing single electrons (``$1e$") are selected without the
vertex and timing requirements described for $\beta\beta$ events. Events
containing a single electron and a certain number of photons (``$1eN\gamma$") do
have similar timing requirements to $\beta\beta$ events in order to eliminate
crossing particles. Events that, by contrast, have timing characteristics
consistent with particles crossing the detector, are used to constrain
``external" backgrounds originating from outside the detector.  Finally,
``$1e1\alpha$" events contain a single electron and delayed $\alpha$ particle
consistent with $^{214}{\rm Bi}-^{214}{\rm Po}$ sequential decays.


The data are compared to simulated signal and background samples. In all cases
the decays are generated using the {\small\tt DECAY0}~\cite{decay0} program and
passed through a detailed {\small\tt GEANT3}~\cite{geant3} based detector
simulation, before being processed with the same reconstruction algorithms and
event selection as the data.

\section{\label{sec:backgrounds}Backgrounds and Control Channels}


Through interactions in or near the source foil, \mbox{$\beta$-decaying}
isotopes or external $\gamma$ rays can give rise to events containing two
electrons. Due to the high \mbox{$Q_{\beta\beta}$ value} of \ca{} very few
isotopes form significant backgrounds in the vicinity of the \bbzero{}
signal. Backgrounds are classified as ``internal" if they originate from the
source foil itself and ``external" if they originate from outside of the
tracking volume. Radon progeny deposited on the source foil or nearby tracker
wires form a third background category.

\begin{figure}[htb]
  \includegraphics[width=0.48\textwidth]{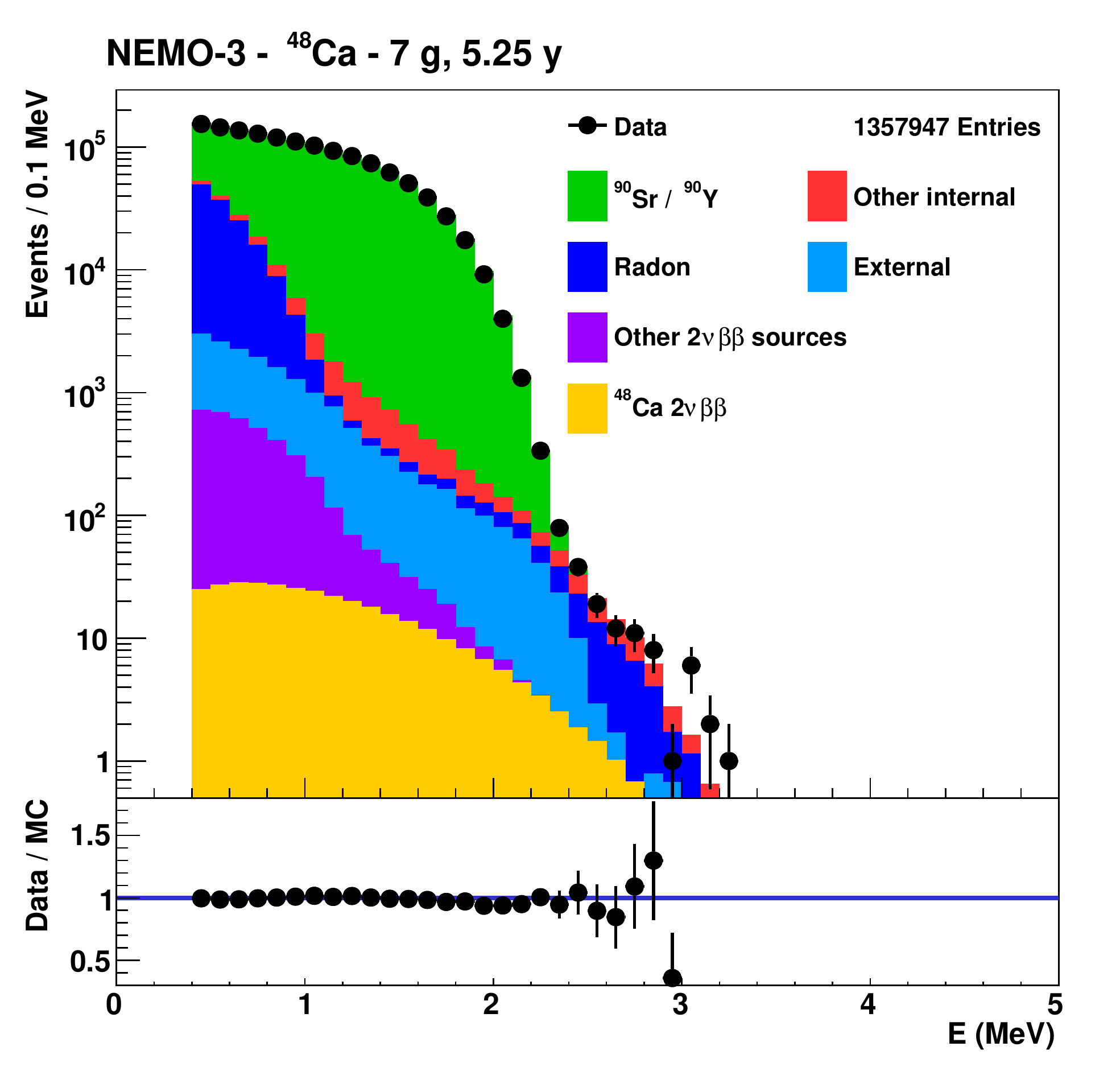}
  \caption{\label{fig:1e} The energy spectrum of electrons in events containing
    a single electron compared to the fitted background contributions (top
    panel) and as a ratio to the total Monte Carlo prediction (bottom
    panel). The dominant contribution is from $^{90}{\rm Sr}/^{90}{\rm Y}$ with
    an end point of $2.3$\,${\rm MeV}$. At higher energies the background from
    the radon daughter $^{214}{\rm Bi}$ becomes significant.  There are
    subleading contributions from other internal backgrounds, external
    backgrounds and contamination of events from neighboring source foils.}
\end{figure}

The \ca{} sources in NEMO-3 suffer from contamination by the chemically similar
element strontium.
Both $^{90}{\rm Y}$ and its parent $^{90}{\rm Sr}$, with which it is in secular
equilibrium, are essentially pure $\beta$ emitters and their activity is
constrained predominantly by the measurement using the $1e$ channel, as shown in
Fig.~\ref{fig:1e}. The measured $^{90}{\rm Sr}/^{90}{\rm Y}$ activity is
observed to decrease with a half-life of $(24.8\pm 0.5)$\,${\rm yr}$ over the
almost 8-yr running period of the experiment. The discrepancy with the
expected $28.8$-${\rm yr}$ half-life~\cite{Sr90} corresponds to differences of
$(1-2)\%$ between the measured and predicted distribution of $^{90}{\rm
  Sr}/^{90}{\rm Y}$ background events across the lifetime of the experiment, due
to imperfect modeling of the time dependence of the detector response. The
resulting systematic uncertainty on the results presented here is negligible.

The external backgrounds indicated in Fig.~\ref{fig:1e} and elsewhere are
modeled as combinations of $^{228}{\rm Ac}/^{212}{\rm Bi}/^{208}{\rm Tl}$,
$^{214}{\rm Bi}$, $^{60}{\rm Co}$ and $^{40}{\rm K}$ in the photomultiplier
tubes, scintillator blocks, iron structure and laboratory air surrounding the
detector~\cite{nemo3_background}. All of these components are allowed to vary
from their central values in a fit to $1e$ and $1e1\gamma$ events that have a
timing signature consistent with a particle of external origin passing through
the \ca{} source. Due to detector inhomogeneities, the various external
background components measured in the analysis of \ca{} data are found to differ 
from their whole-detector averaged levels by between $-15\%$ and $+40\%$.

\begin{figure}[tb]
  \includegraphics[width=0.48\textwidth]{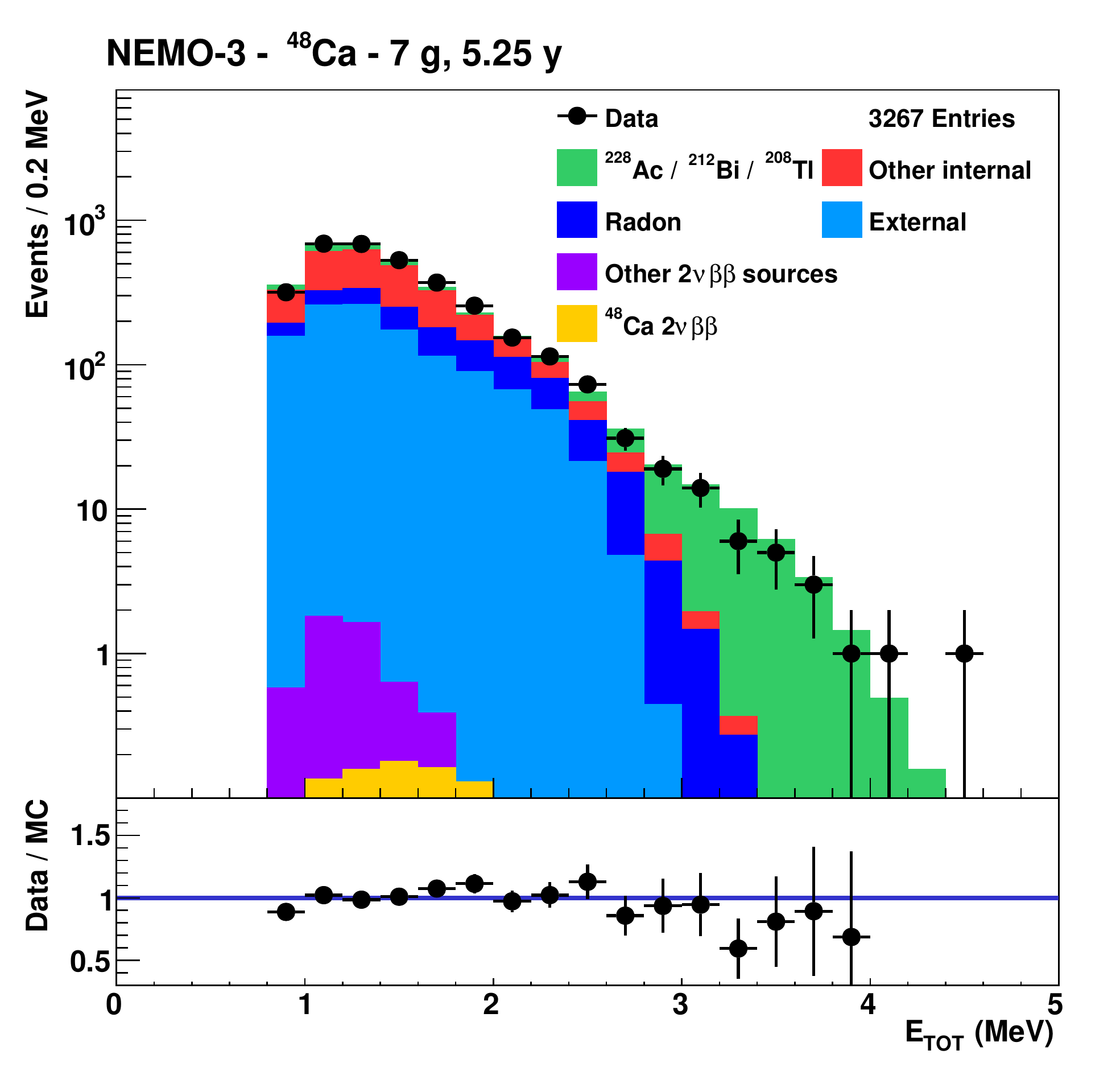}
  \caption{\label{fig:1e1g} The distribution of the total energy measured in
    events with one electron and one photon (top panel) and as a ratio to the
    total Monte Carlo prediction (bottom panel).  The higher end of the
    distribution is dominated by the contribution due to the $\beta$ decay of
    $^{208}$Tl to $^{208}$Pb which always occurs with the emission of one or
    more photons, releasing a total energy of 5 MeV.}
\end{figure}

Events containing a gamma in addition to a single electron provide an additional
control channel to constrain internal backgrounds such as $^{208}{\rm Tl}$,
$^{214}{\rm Bi}$ and $^{152}{\rm Eu}$. Figure~\ref{fig:1e1g} shows the sum of
the electron and photon energies in $1e1\gamma$ events. On top of a significant
external background component, there is a contribution at low energies from
$^{90}{\rm Sr}/^{90}{\rm Y}$ decays in which the electron has undergone
bremsstrahlung, with the other internal backgrounds constrained predominantly in
the region $E_{e+\gamma}\gtrsim 2.0$\,${\rm MeV}$.  The $1e2\gamma$ events
provide an additional statistically limited constraint, particularly on internal
contamination by $^{208}{\rm Tl}$.

\begin{figure}[htb]
  \includegraphics[width=0.48\textwidth]{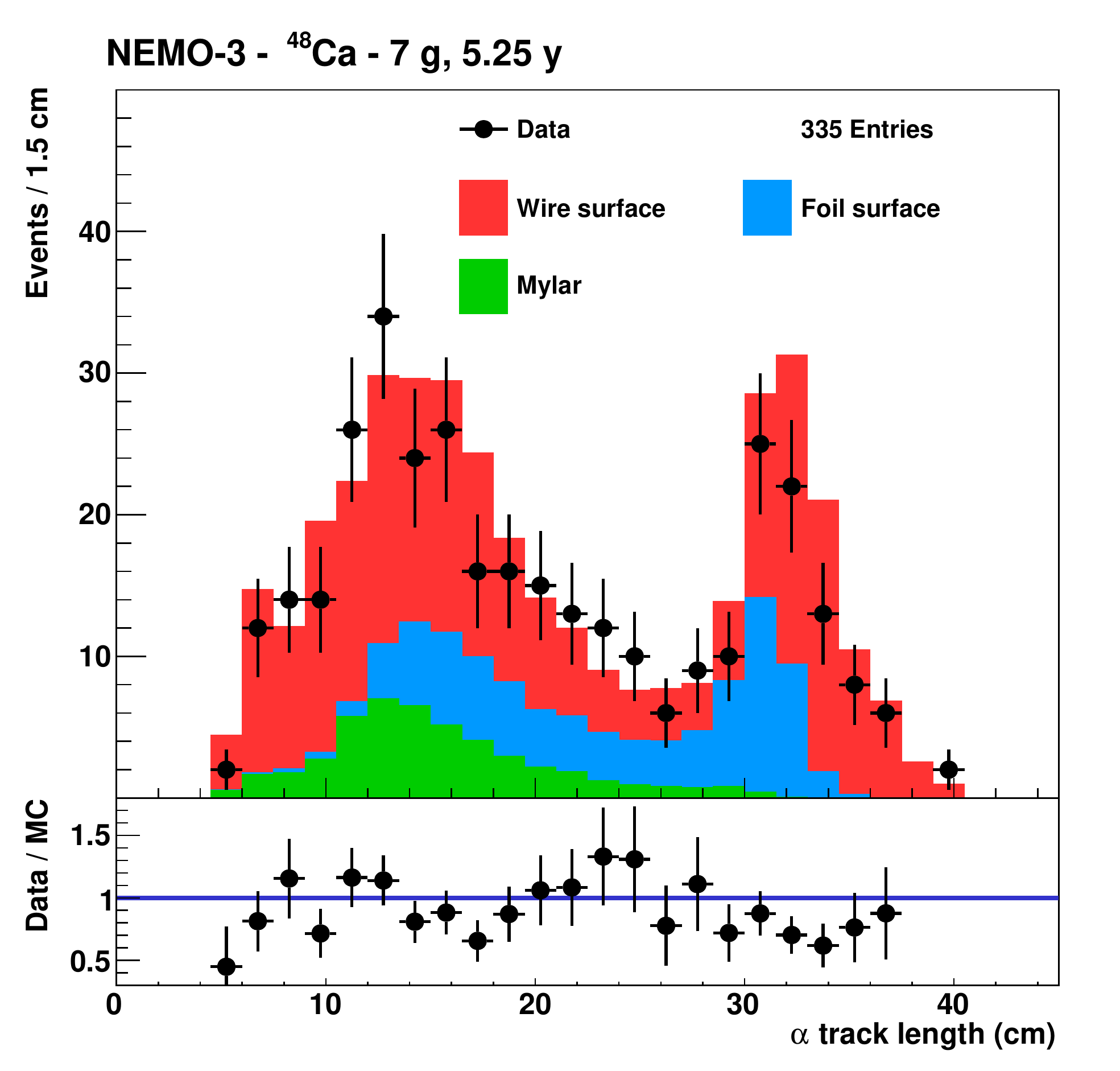}
  \caption{\label{fig:1e1a} The length of delayed tracks in $1e1\alpha$ events,
    showing the contribution from radon progeny deposited on the wires of the
    tracking detector (red) and surface of the source foil (blue). A small
    contribution consistent with having an origin inside the Mylar surrounding
    the ${\rm CaF_{2}}$ disks, peaking at small track lengths, is also
    shown. The ratio of data events to the total Monte Carlo prediction is shown
    in the bottom panel.}
\end{figure}

The final background control sample consists of events containing a single
electron and a delayed track consistent with being an $\alpha$ particle from
$^{214}{\rm Bi}-^{214}{\rm Po}$ coincident decays. The minimum electron energy
requirement is reduced to $200$\,${\rm keV}$ in order to increase statistics,
and $\alpha$ tracks comprise one or more delayed hits in the vicinity of the
electron vertex.  The distribution of the $\alpha$ track length in
Fig.~\ref{fig:1e1a} shows a clear signal for radon progeny deposited on the
surface of tracker wires close to the source foil, as well as a component
deposited on the foil itself. Due to the small efficiency for $\alpha$ particles
generated in the ${\rm CaF_{2}}$ powder to reach the tracking volume, this
channel does not constrain the internal $^{214}{\rm Bi}$ contamination of the
isotope and provides only a weak signal for contamination inside the Mylar
surrounding the source.


High-purity germanium (HPGe) measurements of the \ca{} source disks were made
following the dismantling of the NEMO-3 detector. Activities of $^{152}{\rm Eu}$
and $^{40}{\rm K}$ and upper limits on the activities of $^{234m}{\rm Pa}$ and
$^{228}{\rm Ac}$ are derived and used as Gaussian constraints in the final
background model.

A simultaneous binned log-likelihood fit is performed on all of the control
samples discussed in Sec.~\ref{sec:selection}, as well as the HPGe
constraints. In all cases the HPGe measurements are found not to be in
significant tension with the NEMO-3 data. The resulting internal background
activities are summarized in the first column of Table~\ref{tab:backgrounds}.
\begin{table}[b!]
  \renewcommand{\arraystretch}{1.2}
  \begin{center}
    \begin{ruledtabular}
      \begin{tabular}{l c c c} 

        \multirow{2}{*}{Contribution}                   & \multirow{2}{*}{$A$~(mBq)} &  \multirow{2}{*}{$N$} &  $N$~for                \\
                                                        &                            &                       &  $E_{\rm TOT}>1.8$\,MeV \\ \hline
        $^{90}{\rm Sr}/^{90}{\rm Y}$ (in $\beta\beta$)  & $32.3 \pm 0.1\phantom{0}$  &  $997$                &  $32.1$                 \\
        $^{228}{\rm Ac}/^{212}{\rm Bi}/^{208}{\rm Tl}$  & $0.07 \pm 0.01$            &  $6.8$                &  $1.5$                  \\
        $^{214}{\rm Pb}/^{214}{\rm Bi}$                 & $0.08 \pm 0.01$            &  $12.9$               &  $3.8$                  \\
        $^{152}{\rm Eu}$                                & $\phantom{0} 0.5 \pm 0.1\phantom{0}$ &  $0.13$     &                         \\
        $^{40}{\rm K}$                                  & $0.49 \pm 0.08$            &  $0.27$               &                         \\
        $^{234m}{\rm Pa}$                               & $\phantom{0}0.3   \pm 0.1\phantom{0}$ &  $8.4$     &  $1.6 \times 10^{-2}$   \\
        Non-$^{48}{\rm Ca}$ sources                     &                            &  $16.0$               &  $3.1$                  \\
        External                                        &                            &  $7.6$                &  $0.81$                 \\
        Radon                                           &                            &  $7.9$                &  $2.9$                  \\
        Total background                                &                            &  $1057$               &  $44.2$                 \\
        $^{48}{\rm Ca}$ $2\nu\beta\beta$                & (3.0 $\pm$ 0.3) $\times$ 10$^{-2}$  &  $302$       &  $153$                  \\ \hline
        Data                                            &                            &  $1368$               &  $192$                  \\
      \end{tabular}
    \end{ruledtabular}
    \caption{Summary of the fitted internal background activities ($A$), the
      expected number of background events ($N$) from all sources, the fitted
      $^{48}{\rm Ca}$ $2\nu\beta\beta$ signal and the observed number of
      two-electron events. The numbers of events in the region $E_{\rm
        TOT}>1.8$\,MeV, in which the $2\nu\beta\beta$ contribution has been
      fitted, are also given. Where multiple isotopes are listed on a single
      line then secular equilibrium is assumed. Note that for $^{228}{\rm
        Ac}/^{212}{\rm Bi}/^{208}{\rm Tl}$ the quoted activity is for
      $^{228}{\rm Ac}$ and the activity of $^{208}{\rm Tl}$ will therefore be
      smaller by a branching fraction of $35.9\%$. The uncertainties on the
      quoted activities are statistical only.  }
    \label{tab:backgrounds}
  \end{center}
\end{table}

\section{\label{sec:2vBB}Two-Neutrino Double-Beta Decay}

\begin{figure}[htb]
  \includegraphics[width=0.48\textwidth]{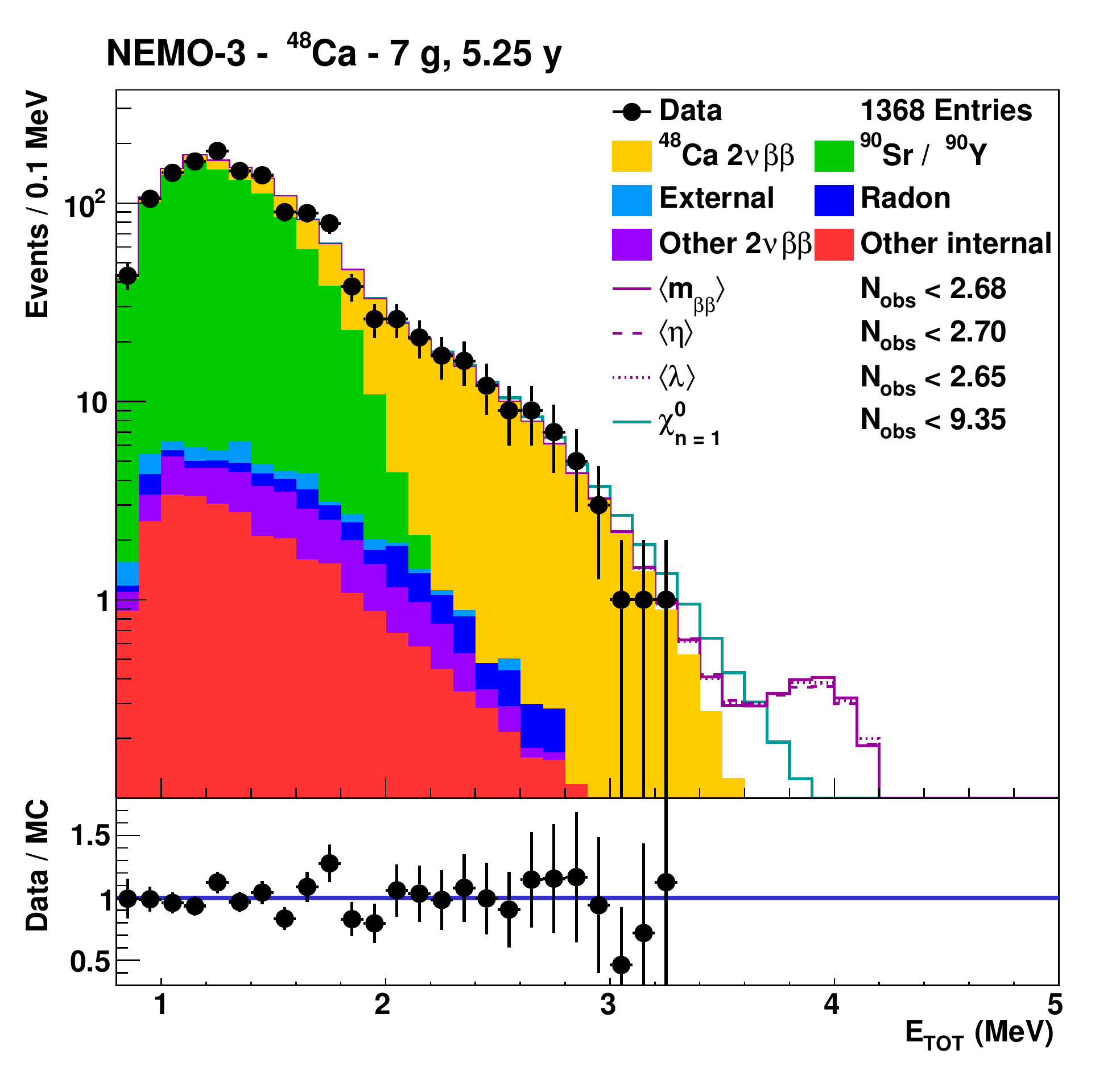}
  \caption{\label{fig:2e_sumE} The distribution of the summed electron energy in
    two-electron events. At low energies $^{90}{\rm Sr}/^{90}{\rm Y}$ background
    events dominate, while other internal, external, radon and non-\ca{}
    backgrounds are relatively small.  The $2\nu\beta\beta$ signal is clearly
    visible at higher energies. Four open histograms represent limits on
    non-standard model double-beta decay processes, with $90\%$ confidence
    level upper limits on the corresponding event yields $N_{\rm obs}$ also
    given. The ratio of data events to the total Monte Carlo prediction is shown
    in the bottom panel.}
\end{figure}

Events containing exactly two electrons are selected with the requirements
listed in Sec.~\ref{sec:selection}. The distribution of $E_{\rm
  TOT}=E_{1}+E_{2}$ is shown in Fig.~\ref{fig:2e_sumE} along with the fitted
backgrounds.
The $^{90}{\rm Sr}/^{90}{\rm Y}$ decays can give rise to two-electron events,
for example through M\o{}ller scattering. The yield of events fitted in the
$\beta\beta$ sample alone and shown in Fig.~\ref{fig:2e_sumE} results in an
activity for $^{90}{\rm Sr}/^{90}{\rm Y}$ that is $(9\pm4)\%$ higher than that
measured in the single-electron control sample. Although not significant, this
difference is considered as a source of systematic uncertainty.

The $2\nu\beta\beta$ event yield is fitted in the high-energy region of the
$E_{\rm TOT}$ distribution. An optimization of the total (statistical plus
systematic) uncertainty indicates an optimum requirement of $E_{\rm
  TOT}>1.8$\,${\rm MeV}$. Table~\ref{tab:backgrounds} gives the fitted and
observed event yield over the full range $E_{\rm TOT}>0.8$\,${\rm MeV}$ and in
the signal fit range $E_{\rm TOT}>1.8$\,${\rm MeV}$. In the fit range the number
of signal events divided by the square root of the number of background events,
$S/\sqrt{B}\approx 23$.  The isotopes $^{90}{\rm Sr}/^{90}{\rm Y}$ are the most
important background.  The fitted number of events $N=153$ and the selection
efficiency calculated from Monte Carlo simulation, $\epsilon=3.1\%$, correspond
to a half-life given by:
\begin{displaymath}
T_{1/2} = \frac{\epsilon}{N}\cdot \frac{N_{A}m}{M}\cdot \ln(2)\cdot T \; ,
\end{displaymath}
where $N_{A}$ is Avogadro's number, $m$ is the mass of \ca{} with molar mass
$M$, and $T$ is the total exposure time.

In addition to the statistical uncertainty on the fitted number of signal
events, the measurement of the two-neutrino half-life is subject to a number of
systematic uncertainties as listed in Table~\ref{tab:systematics}. The majority
of the contributions relate to uncertainties in the modeling of signal and
backgrounds.  The discrepancy between the $^{90}{\rm Sr}/^{90}{\rm Y}$ activity
measured in the $1e$ and $2e$ samples is likely due to imperfect modeling of
the mechanisms by which single $\beta$ electrons give rise to two-electron
event signatures in the NEMO-3 detector. The central value for the \ca{}
$2\nu\beta\beta$ half-life is obtained by fitting the
$^{90}{\rm Sr}/^{90}{\rm Y}$ activity simultaneously with the signal in the
$2e$ sample; constraining it to the activity measured in the $1e$ sample gives
rise to a change between $-2.4\%$ and $+2.0\%$ in $T^{2\nu}_{1/2}$.
%
%
Uncertainties in external and radon backgrounds give rise to changes in the
measured half-life approximately one order of magnitude smaller than this.

%
%
The fractional uncertainty on the \ca{} enrichment fraction of $2.1\%$
translates directly into a corresponding uncertainty on the measured
two-neutrino half-life. Systematic uncertainties are also attributed to the
source construction (diameter, thickness and material composition) as well as
the precision ($\approx 1^{\circ}$) with which the orientation of the \ca{}
source strips is known within the NEMO-3 detector. Altogether these source
construction uncertainties combine to give an asymmetric $+3.7\%$ and $-5.5\%$
uncertainty on the extracted \bbtwo{} half-life.

%
%
The largest systematic uncertainties relate to the detector response and
calibration.  A comparison of data and Monte Carlo simulations for dedicated
$^{207}{\rm Bi}$ calibration runs indicates an uncertainty on the selection
efficiency for double-beta decay topologies of $\approx
7\%$~\cite{nemo3_mo100}. A calorimeter energy scale uncertainty of $1\%$
determined during dedicated scans of the calorimeter optical modules prior to
construction and verified using \emph{in situ} calibration sources translates into a
$4.2\%$ uncertainty on the \bbtwo{} activity. One of the most difficult
distributions to simulate accurately is the opening angle of the two electrons
in double-beta-like event topologies, which requires accurate modeling not
only of electron scattering in the source material but also the performance of
the reconstruction algorithms for tracks with varying spatial
separations. Separating two-electron events into exclusive samples comprising
events with both electrons on the same side (SS) or opposite side (OS) of the
source foil gives rise to fitted activities that are different from the combined
sample by $-14\%$ (SS) and $+11\%$ (OS) respectively. Although this discrepancy
has a statistical significance of only $1.2\sigma$, it is included here as an
asymmetric systematic uncertainty on the half-life obtained from the inclusive
$2e$ sample.

For the source mass and exposure stated in Sec.~\ref{sec:NEMO-3}, the
resulting half-life for the \bbtwo{} decay of \ca{} is:
\begin{displaymath}
T^{2\nu}_{1/2} \,=\, [6.4\, ^{+0.7}_{-0.6}{\rm (stat.)} \, ^{+1.2}_{-0.9}{\rm
    (syst.)}] \times 10^{19}\,{\rm yr}\; ,
\end{displaymath}
where the systematic uncertainty is the sum in quadrature separately for upward
and downward changes in the half-life of all the systematics described above and
summarized in Table~\ref{tab:systematics}.
\begin{table}[b!]
  \renewcommand{\arraystretch}{1.2}
  \begin{center}
    \begin{ruledtabular}
      \begin{tabular}{l c} 
        Origin                                  & Uncertainty on  $T^{2\nu}_{1/2}$ \\   \hline 
        $^{90}{\rm Sr}/^{90}{\rm Y}$ background & $[+2.0, -2.4] \%$                \\
        Other backgrounds                       & $\pm 0.3 \%$                     \\
        \ca{} enrichment fraction               & $\pm 2.1 \%$                     \\      
        \ca{} source construction               & $[+3.7, -5.5] \%$                \\      
        Electron reconstruction efficiency      & $[+7.5, -6.5] \%$                \\      
        Calorimeter energy scale                & $[+4.4, -4.0] \%$                \\      
        Two-electron angular distribution       & $[+16, -10] \%$                  \\
        Total                                   & $[+19, -14] \%$                  \\      
      \end{tabular}
    \end{ruledtabular}
    \caption{Systematic uncertainties on the measured half-life for the \bbtwo{}
      decay of \ca{}.}
    \label{tab:systematics}
  \end{center}
\end{table}

\section{\label{sec:0vBB}Neutrinoless Double-Beta Decay}
A search for neutrinoless modes of double-beta decay is performed on events
selected with the criteria outlined in Sec.~\ref{sec:selection}, using the
measurements of backgrounds and the~\bbtwo{} signal strength given in
Secs.~\ref{sec:backgrounds} and~\ref{sec:2vBB}, respectively. The main
background for~\bbzero{} searches is the \bbtwo{} signal, which overlaps with
the~\bbzero{} peak given the finite resolution of the calorimeter.

Four~\bbzero{} mechanisms are investigated in this work, where the decay
proceeds via the exchange of light neutrinos, which largely shares its
kinematics with \mbox{$R$-parity} violating ($\slashed{R}_p$) supersymmetric
processes~\cite{vergados_2012}, through right-handed currents coupling
right-handed quarks to right-handed leptons ($\lambda$) and left-handed quarks
to right-handed leptons ($\eta$)~\cite{supernemo_new_physics}, and with the
emission of a single Majoron with spectral index $n=1$~\cite{bamert_1995}.

Limits on the~\bbzero{} processes are obtained using a modified frequentist
method based on a binned log-likelihood ratio test statistic
(CL$_s$)~\cite{cls}. The statistic is calculated over the entire energy range
above $0.8$\,${\rm MeV}$, with the background-only hypothesis providing a good
fit to the data with an observed $p$ value (1-CL$_b$) of $0.83$.

The region above $3.4$\,${\rm MeV}$ has the highest sensitivity to the~\bbzero{}
peak. No events are observed in this window, illustrating the benefit of the
high $Q_{\beta\beta}$ value of~\ca{}, and a simple counting experiment limit
obtained in this window is comparable to that obtained using the CL$_s$ method
over the larger energy range.  Requiring a minimum energy of $3.6$\,${\rm MeV}$
reduces the signal efficiency by a modest $15\%$ while reducing the background
expectation, dominated by~\bbtwo{} events, to $\approx 0.1$~events. This
suggests the feasibility of background-free measurements with at least one order
of magnitude longer exposures using the NEMO-3 technique.

When calculating the test statistic, the background and signal distributions are
scaled by random factors drawn from Gaussian distributions with widths
reflecting the uncertainties on the background normalization and signal
efficiency. An estimate of the effect of systematic uncertainties on the limit
is thus included in the resulting distributions of the test statistic from which
the confidence intervals are extracted. The estimate of the systematic
uncertainty on the background model given in Sec.~\ref{sec:2vBB} is used, and
the statistical uncertainty on the magnitude of the~\bbtwo{} signal given in the
same section is applied to that contribution. Both these uncertainties have
negligible effects on the extracted limits, with the exception of the 
Majoron-emission mode. The largest effect arises from the uncertainty on the signal
efficiency. Given that the dominant contribution to the uncertainty on
the~\bbtwo{} efficiency is estimated from the data, which is not possible for
the~\bbzero{} case, the uncertainty on the latter signal efficiency is assumed
to be equal in magnitude to the former.

The combined effect of the uncertainties outlined above on the limit placed on
the light neutrino exchange~\bbzero{} process does not exceed 5\%.  The results
of the limit-setting procedure are given in Table~\ref{tab:0nu_lims} and shown
as open histograms in Fig.~\ref{fig:2e_sumE}.

\begin{widetext}

  \onecolumngrid

  \begin{table}[h!]
    \begin{ruledtabular}
      \begin{tabular}{l c c c c c }
        \multirow{2}{*}{Mechanism}                                & Efficiency & \multicolumn{2}{c}{$N^{\rm lim}$ 90\% C.L.}  &   \multicolumn{2}{c}{$\; T^{0\nu}_{1/2}$ 90\% C.L. (10$^{22}$ yr) } \\
                                                                  & (\%)       &   Expected          & Observed      &   Expected         &    Observed \\ \hline
        Light neutrino exchange $/$ $\slashed{R}_p$ supersymmetry & 16.9       &    $<$ 2.67 -- 3.15 & $<$ 2.68      &   $>$ 1.71 -- 2.02 &    $>$ 2.02 \\
        \multirow{2}{*}{Right-handed currents} ~\, $\eta$         & 15.8       &    $<$ 2.69 -- 3.21 & $<$ 2.70      &   $>$ 1.57 -- 1.88 &    $>$ 1.87 \\
        ~~~~~~~~~~~~~~~~~~~~~~~~~~~~~~~\,\, $\lambda$             & 9.91       &    $<$ 2.65 -- 3.07 & $<$ 2.65      &   $>$ 1.03 -- 1.20 &    $>$ 1.19 \\
        Single Majoron emission $(n=1)$                           & 13.4       &    $<$ 8.54 -- 17.3 & $<$ 9.35      &   $>$ 0.25 -- 0.50 &    $>$ 0.46 \\
      \end{tabular}
    \end{ruledtabular}
    \caption{Limits on~\bbzero{} modes in \ca{} for $E_{\rm TOT}>0.8$\,${\rm
        MeV}$. The signal efficiencies and 90\% C.L. upper limits on the number
      of events $(N^{\rm lim})$ are shown for the four investigated decay
      mechanisms. The limits expected given the background-only hypothesis are
      shown, with the ranges representing one standard deviation fluctuations of
      the background model and the systematic uncertainty on the signal
      efficiency.}
    \label{tab:0nu_lims}
  \end{table}
\end{widetext}

\twocolumngrid

The most stringent limit is obtained for the light Majorana neutrino exchange
mechanism, where fewer than $2.68$ events are observed at 90\% C.L. With a
signal detection efficiency of 16.9\%, this corresponds to a lower limit of
\mbox{$2.0 \times 10^{22}\,{\rm yr}$} on the~\bbzero{} half-life. This result is
within the range expected for the background-only hypothesis. As noted in the
Introduction, a search for the same process with scintillating CaF$_2$ crystals
has achieved a better limit of \mbox{$5.8 \times 10^{22}\,{\rm yr}$} with
roughly half of the exposure and a significantly higher
efficiency~\cite{elegant_2008}.
        
The \bbzero{} half-life lower limit can be converted into an upper limit on the
effective Majorana neutrino mass $\braket{m_{\beta\beta}}$.
Using the axial-vector coupling constant \mbox{$g_A = 1.27$}, the phase space
factor from~\cite{phase_space} and NMEs calculated in the shell-model
framework~\cite{nme_ism_ca48, horoi_2013, iwata_2016} gives $\braket{m_{\beta\beta}} < 12 -
24$ eV, whereas extending the NME selection to include calculations in the
QRPA~\cite{suhonen_1993,simkovic_2013,nme_qrpa_ca48}, interacting boson 
model~\cite{nme_ibm_ca48_2015} and energy density functional~\cite{nme_edf_ca48} 
frameworks yields a wider range:
\mbox{$\braket{m_{\beta\beta}} < 6.0 - 26$ eV.}

In the context of supersymmetry, the same limit can be used to extract an upper
limit on the $\slashed{R}_p$ coupling constant $\lambda'_{111}$, assuming the
decay proceeds via gluino exchange~\cite{vergados_2012}. With the matrix
elements given in~\cite{horoi_2013, wodecki_1999} a bound of $\lambda'_{111} <
(0.11 - 0.14) \times f$ is obtained on the coupling constant, with $f =
\left(\frac{m_{\tilde{q}}}{1 {\rm TeV}}\right)^2 \left(\frac{m_{\tilde{g}}}{1
  {\rm TeV}}\right)^{1/2}$, where $m_{\tilde{q}}$ and $m_{\tilde{g}}$ are the
squark and gluino masses, respectively.

The existence of right-handed weak currents would lead to~\bbzero{} decays with
different electron kinematics from the neutrino exchange mechanism. The summed
energy spectra do not differ significantly from the light neutrino exchange
mode, giving consistent limits on the number of observed events. However, the
right-handed modes have reduced detection efficiencies, particularly in the
$\lambda$ case, where there is a significant asymmetry in the energies of the
two electrons. The resulting lower limits on the half-lives of these processes
are $1.9 \times 10^{22}\,{\rm yr}$ and $1.2 \times 10^{22}\,{\rm yr}$ at 90\%
C.L. for the $\eta$ and $\lambda$ modes, respectively. These half-lives
translate into upper bounds on the coupling constants of $\braket{\eta} < (0.74
- 54) \times 10^{-7}$ and $\braket{\lambda} < (7.6 - 47) \times 10^{-6}$, using
phase space factors and matrix elements from~\cite{suhonen_1998}. A unique
strength of NEMO-like experiments is the ability to measure the kinematics of
the individual electrons, which would allow for the distinction between
right-handed currents and light neutrino exchange as the underlying mechanisms
of~\bbzero{} in the event of a positive
observation~\cite{supernemo_new_physics}.

Lastly, a search is made for~\bbzero{} accompanied by the emission of a single
Majoron. Models considered in the literature predict~\bbzero{} to occur with the
emission of one or two Majorons, with the phase space factor $G$ governed by $G
\propto (Q_{\beta\beta} - E_{\rm TOT})^n$, where $n$ is the spectral index,
commonly used to categorize decays arising from different
models~\cite{bamert_1995}. Decays with larger $n$ (e.g. those with the emission
of two Majorons) have broader summed energy spectra peaked at lower values and
are thus more difficult to separate from the~\bbtwo{} signal and other
backgrounds. Given the relatively low statistics and high level of background at
the lower end of the spectrum studied in this work, only the $n=1$ case is
considered. The upper limit on the number of observed events is $9.35$ at 90\%
C.L., which is combined with a detection efficiency of 13.4\% to give a lower
half-life limit of $4.6 \times 10^{21}\,{\rm yr}$. This result improves on the
previously published best limit of $7.2 \times 10^{20}\,{\rm
  yr}$~\cite{barabash_ca48_majoron}, obtained with data from~\cite{bardin_1970}.
An upper limit on the coupling between the $\nu_e$ and the Majoron of
$\braket{g_{ee}} < (1.0 - 4.3) \times 10^{-4}$ is extracted from the half-life
limit using the phase space factor from~\cite{suhonen_1998} and matrix elements
from~\cite{nme_ibm_ca48_2015, nme_ism_ca48, nme_edf_ca48, suhonen_1993,
  horoi_2013, simkovic_2013, iwata_2016}.

\section{\label{sec:summary}Summary and Conclusions}

The NEMO-3 experiment has investigated the double-beta decay of \ca{} with a
small source comprising just $7$\,${\rm g}$ of isotope but with a total exposure
time of more than five years. A larger and purer sample of $\beta\beta$ events
has been selected than in previous experiments and the half-life for the
standard model two-neutrino double-beta decay mode has been measured to be
\mbox{$T^{2\nu}_{1/2}\,=\,[6.4\, ^{+0.7}_{-0.6}{\rm (stat.)} \,
    ^{+1.2}_{-0.9}{\rm (syst.)}] \times 10^{19}\,{\rm yr}$}.  This measured
half-life is consistent with previous experimental measurements but has
significantly smaller uncertainties.
Note that this result differs from and supercedes a preliminary measurement of
the \bbtwo{} half-life for \ca{} previously reported by the NEMO-3
Collaboration~\cite{2vBB_review, barabash_2015}. The change in reported
half-life is due primarily to a modified description of the source geometry
following the decommissioning of the NEMO-3 detector.
The half-life reported in this work is longer than suggested by shell-model
calculations, although the significance of this discrepancy is only at the level
of $2\sigma$.

A search for \bbzero{} decays of \ca{} has been performed in the same
data set. No signal has been found, and a lower limit on the half-life for the
light Majorana neutrino exchange mechanism for \bbzero{} has been determined to
be \mbox{$2.0 \times 10^{22}\,{\rm yr}$} at 90\% C.L. Limits have also been
placed on $\slashed{R}_p$ supersymmetry, right-handed currents and
Majoron-emission models.

As expected for the high-$Q_{\beta\beta}$ isotope \ca{}, the region of interest
for \bbzero{} decays has a very low level of residual background arising almost
entirely from the high-energy tail of the \bbtwo{} decay mode. Further
investigation of the double-beta decay of \ca{} using a similar experimental
technique but with larger exposures would therefore lead to improved limits on
\bbzero{} as well as provide a more precise measurement of the \bbtwo{}
half-life.  However, progress in the enrichment of \ca{} will be required for
this to be experimentally feasible.

\begin{acknowledgments}
We thank the staff of the Modane Underground Laboratory for their technical
assistance in running the experiment. We acknowledge support by the grants
agencies of the Czech Republic, CNRS/IN2P3 in France, RFBR in Russia, STFC in
the U.K. and NSF in the U.S.
\end{acknowledgments}


\begin{thebibliography}{6}

\bibitem{0vBB_review} S.~Pascoli, S.~T.~Petcov and T.~Schwetz, Nucl.~Phys.~B
  {\bf 734} (2006) 24.

\bibitem{2vBB_review} R.~Saakyan, Ann.~Rev.~Nucl.~Part.~Sci. {\bf 63} (2013)
  503.

\bibitem{barabash_2015} A.~S.~Barabash, Nucl.~Phys.~A {\bf 935} (2015) 52.

\bibitem{cuoricino} K.~Alfonso {\it et al.}, Phys.\ Rev.\ Lett. {\bf 115} (2015)
  102502.
  
\bibitem{kamland_zen} A.~Gando {\em et al.}, Phys.~Rev.~Lett. {\bf 110} (2013)
  062502.

\bibitem{gerda} M.~Agostini {\em et al.}, Phys.~Rev.~Lett. {\bf 111} (2013)
  122503.

\bibitem{nemo3_mo100} R.~Arnold {\it et al.}, Phys.\ Rev.\ D {\bf 89} (2014)
  111101.

\bibitem{exo200} J.~B.~Albert {\em et al.}, Nature {\bf 510} (2014) 229.

\bibitem{ca48_Qbb} A.~A.~Kwiatkowski {\em et al.}, Phys.~Rev.~C {\bf 89} (2014)
  045502.
  
\bibitem{nme_qrpa_ca48} J.~Suhonen, AIP Conf.~Proc. {\bf 1488} (2012) 326.

\bibitem{nme_ism_ca48} E.~Caurier {\em et al.}, Phys.~Rev.~Lett. {\bf 100}
  (2008) 052503.

\bibitem{iwata_2016} Y.~Iwata {\em et al.}, Phys.~Rev.~Lett. {\bf 116} (2016)
  112502.
  
\bibitem{nme_edf_ca48} T.~R.~Rodriguez and G.~Martinez-Pinedo,
  Phys.~Rev.~Lett. {\bf 105} (2010) 252503.

\bibitem{suhonen_1993} J.~Suhonen, J.~Phys.~G {\bf 19} (1993) 139.
\bibitem{simkovic_2013} F.~\v{S}imkovic {\em et al.}, Phys.~Rev.~C {\bf 87}
  (2013) 045501.
  
\bibitem{nme_ibm_ca48_2015} J.~Barea, J.~Kotila and F.~Iachello, Phys.~Rev.~C
  {\bf 91} (2015) 034304.
  
\bibitem{horoi_2013} M.~Horoi, Phys.~Rev.~C {\bf 87} (2013) 014320.

\bibitem{balysh_1996} A.~Balysh {\em et al.}, Phys.~Rev.~Lett. {\bf 77} (1996)
  5186.

\bibitem{brudanin_2000} V.~B.~Brudanin {\em et al.}, Phys.~Lett.~B {\bf 495}
  (2000) 63.


\bibitem{caurier_1994} E.~Caurier {\em et al.}, Phys.~Rev.~C {\bf 50} (1994)
  225.

\bibitem{caurier_2012} E.~Caurier, F.~Nowacki and A.~Poves, Phys.~Lett.~B {\bf
  711} (2012) 62.

\bibitem{horoi_2007} M.~Horoi, S.~Stoica and B.~A.~Brown, Phys.\ Rev.\ C {\bf
  75} (2007) 034303.

\bibitem{zhao_1990} L.~Zhao, B.~A.~Brown and W.~A.~Richter, Phys.\ Rev.\ C {\bf
  42} (1990) 1120.


  
\bibitem{raduta_2011} C.~M.~Raduta, A.~A.~Raduta and I.~I.~Ursu, Phys.\ Rev.\ C
  {\bf 84} (2011) 064322.


\bibitem{mccarthy_1955} J.~McCarthy, Phys.~Rev. {\bf 97} (1955) 1234.

\bibitem{mateosian_1966} E.~der~Mateosian and M.~Goldhaber, Phys.~Rev. {\bf 146}
  (1966) 810.

\bibitem{beijing} K.~You {\em et al.}, Phys.~Lett.~B {\bf 265} (1991) 53.

\bibitem{elegant_2008} S.~Umehara {\em et al.}, Phys.~Rev.~C {\bf 78} (2008)
  058501.

\bibitem{nemo3_detector} R.~Arnold {\em et al.}, Nucl.~Instrum.~Meth.~A {\bf
  536} (2005) 79.


%
\bibitem{nemo3_Nd150_2009} J.~Argyriades {\it et al.}, Phys.~Rev.~C {\bf 80}
  (2009) 032501.
%
\bibitem{nemo3_Zr96_2010} J.~Argyriades {\it et al.}, Nucl.~Phys.~A {\bf 847}
  (2010) 168.
%
\bibitem{nemo3_Te130_2011} R.~Arnold {\it et al.}, Phys.~Rev.~Lett. {\bf 107}
  (2011) 062504.


\bibitem{decay0} O.~A.~Ponkratenko, V.~I.~Tretyak and Y.~G.~Zdesenko,
  Phys.\ Atom.\ Nucl.\ {\bf 63} (2000) 1282.

\bibitem{geant3} {\small\tt GEANT} Detector Description and Simulation Tool,
  CERN Program Library Long Writeup W5013 (1995).

\bibitem{Sr90} E.~Browne, Nuclear Data Sheets 82 (1997) 379.

\bibitem{nemo3_background} J.~Argyriades {\it et al.}, Nucl.\ Instrum.\ Meth.\ A
  {\bf 606} (2009) 449.

%
\bibitem{vergados_2012} J.~D.~Vergados, H.~Ejiri and F.~\v{S}imkovic,
  Rep.~Prog.~Phys. {\bf 75} (2012) 106301.
\bibitem{supernemo_new_physics} R.~Arnold {\it et al.}, Eur.\ Phys.\ J.\ C {\bf
  70} (2010) 927.
\bibitem{bamert_1995} P.~Bamert, C.~P.~Burgess and R.~N.~Mohapatra,
  Nucl.~Phys.~B {\bf 449} (1995) 25.

\bibitem{cls} T.~Junk, Nucl.\ Inst.\ Meth.\ A {\bf 434} (1999) 435.

\bibitem{phase_space} J.~Kotila and F.~Iachello, Phys.~Rev.~C {\bf 85} (2012)
  034316.

\bibitem{wodecki_1999} A.~Wodecki, W.~A.~Kami\'nski and F.~\v{S}imkovic,
  Phys.~Rev.~D {\bf 60} (1999) 115007.

\bibitem{suhonen_1998} J.~Suhonen and O.~Civitarese, Phys.~Rept. {\bf 300}
  (1998) 123.
  
\bibitem{barabash_ca48_majoron} A.~S.~Barabash, Phys.~Lett.~B {\bf 216} (1989)
  257.
\bibitem{bardin_1970} R.~K.~Bardin {\em et al.}, Nucl.~Phys.~A {\bf 158} (1970)
  337.
  
\end{thebibliography}
\end{document}